\newcount\subeqnno
\def\bse{\begingroup
\refstepcounter{equation}
\subeqnno=\arabic{equation}
\setcounter{equation}{0}
\def\theequation{\the\subeqnno\alph{equation}}
}
\def\ese{\setcounter{equation}{\the\subeqnno}\endgroup}

\newcommand{\uu}[1]{\verb!#1!\endgroup}

\newcommand{\mb}[1]{\ifmmode#1\else\mbox{$#1$}\fi}

\newcommand\al{\mb{\alpha}}

\newcommand\be{\mb{\beta}}
\newcommand\ga{\mb{\gamma}}
\newcommand\de{\mb{\delta}}
\newcommand\ep{\mb{\epsilon}}

\newcommand\et{\mb{\eta}}

\newcommand\ka{\mb{\kappa}}
\newcommand\la{\mb{\lambda}}

\newcommand\si{\mb{\sigma}}

\newcommand\ph{\mb{\phi}}

\newcommand\om{\mb{\omega}}

\newcommand\De{\mb{\Delta}}

\newcommand\Ph{\mb{\Phi}}

\newcommand\calC{\mb{{\cal C}}}

\newcommand\calG{\mb{{\cal G}}}
\newcommand\calH{\mb{{\cal H}}}

\newcommand\calL{\mb{{\cal L}}}
\newcommand\calM{\mb{{\cal M}}}

\newcommand\calV{\mb{{\cal V}}}

\newcommand{\m}{{\calM}}

\newcommand{\x}{\mb{\times}}

\newcommand{\cd}{\mb{\partial}}

\newcommand{\beq}{\begin{equation}}
\newcommand{\eeq}{\end{equation}}
\newcommand{\nn}{\nonumber}

\newcommand{\bea}{\begin{eqnarray}}
\newcommand{\eea}{\end{eqnarray}}
 
\newcommand{\fn}{\footnote}
\newcommand{\norm}[1]{\parallel \! {#1} \! \parallel}

\newcommand{\emb}{\mb{{\rm emb}}}
\newcommand{\Ad}{\mb{{\rm Ad}}}

\newcommand{\tr}{\mb{{\rm Tr}}}
\newcommand{\deriv}[2]{\frac{d \hspace{.025em} {#1}}{d {#2}}}
\newcommand{\pderiv}[2]{\frac{\cd {#1}}{\cd {#2}}}
\newcommand{\inprod}[2]{\langle {#1}, {#2} \rangle}
\newcommand{\inproda}[2]{\{ {#1}, {#2} \}}

\documentstyle[12pt]{article}
\topmargin = - 0.5 cm
\begin{document}
\bibliographystyle{unsrt}



\begin{flushright}
CERN-TH/95-210 \\
DAMTP-95-44
\end{flushright}
\begin{center}
\LARGE{EXAMPLES OF EMBEDDED DEFECTS \\
(IN PARTICLE PHYSICS AND CONDENSED MATTER)\\}
\vspace*{0.5cm}
\large{Nathan\  F.\  Lepora
\fn{e-mail: N.F.Lepora@amtp.cam.ac.uk}$^{,1}$
and Anne-Christine\ Davis
\fn{e-mail: A.C.Davis@amtp.cam.ac.uk}$^{,1,2}$\\}
\vspace*{0.2cm}
{\small\em 1) Department of Applied Mathematics and Theoretical
Physics,\\
University of Cambridge, Silver Street,\\
Cambridge, CB3 9EW, U.\ K.\\ }
\vspace*{0.2cm}
{\small\em 2) Theory Division, CERN,\\
Geneva 23, CH-1211, Switzerland.\\} 
\vspace*{0.2cm}

{July 1995}
\end{center}

\begin{abstract}
We present a series of examples designed to clarify the formalism of
the companion paper `Embedded Vortices'. After summarising this
formalism in a prescriptive sense, we run through several examples:
firstly, deriving the embedded defect spectrum for Weinberg-Salam
theory, then discussing several examples designed to illustrate facets
of the formalism. We then  
calculate the embedded defect spectrum for three physical Grand
Unified Theories and conclude with a discussion of vortices formed in
the superfluid $^3$He-A phase transition.
\end{abstract}

\thispagestyle{empty}
\newpage
\setcounter{page}{1}


\section{Introduction.}
\label{sec-1}

Embedded defects have received an impressive amount of interest over
the last couple of years. Principally this is because the Z-string, of
Weinberg-Salam theory, was recently realised to be stable for
part of the parameter space~\cite{Vach92-1}, although it proves to be
unstable in the physical regime~\cite{Jame92, Perk93}; though
there may be other stabilising effects~\cite{Hol92, Vach92-3}. 
However, be it stable or unstable, it may still have important
cosmological consequences --- as indicated by its connection to baryon
number violation~\cite{Vach94-1}.  

The standard model also admits a one-parameter family of
unstable, gauge equivalent vortices: the W-strings~\cite{Vach92-2}.
Together, with the $Z$-string, these constitute a very non-trivial
spectrum of vortices arising from the vacuum structure of the
Weinberg-Salam theory: two gauge-inequivalent families of vortices,
with one family {\em invariant} under the residual electromagnetic
gauge group and the other a one parameter family of gauge
equivalent vortices. Furthermore, only one of these families
has the potential to be stable.

Embedded defects have also been specifically
studied in another symmetry breaking scheme: the GUT
flipped-$SU(5)$~\cite{me1}. One finds an eleven parameter family of gauge
equivalent, unstable vortices plus another globally gauge invariant,
potentially stable vortex (the V-string). 

The general formalism for describing embedded defects was derived by
Vachaspati, {\em et. al.} \cite{Vach94}. Here was described the 
construction of embedded defect solutions for general Yang-Mills
theories: one defines a suitable {\em embedded} subtheory of the
Yang-Mills theory upon which a topological defect solution may be
defined. In extending the embedded subtheory back to the full theory
one loses the stabilising topological nature of the defect, but retains
it as solution to the theory. 

In a companion paper to this~\cite{me2}, the underlying
group theory behind the formalism of~\cite{Vach94} is exploited
to determine the properties and spectrum of embedded defects. 
The purpose of this paper is to provide a list of several examples to
illustrate the formalism of that companion paper.
Firstly, however, we summarise the formalism derived in~\cite{me2}, so
that it may be used prescriptively to determine the spectrum and
stability of embedded defects.  


\section{Summary of Formalism.}
\label{sec-2}

For a Yang-Mills theory, the embedded defects are determined by the
symmetry breaking $G \rightarrow H$. The symmetry breaking depends
upon a scalar field $\Ph$, lying in a vector space $\calV$,
acted on by the $D$-representation of $G$. 
Denoting the Lie algebra of $G$ by $\calG$, the natural action
of $\calG$ upon $\Phi$ is by the derived representation $d$, defined
by $D(e^X) = e^{d(X)}$.

The natural $Gl(\calV)$ invariant inner product on $\calV$ is the real
form  
\beq
\inprod{\Phi}{\Psi} = {\rm Re}(\Phi^\dagger \Psi), \ \ \Phi, \Psi
\in \calV.
\eeq

The general $G$-invariant inner product on $\calG$ is defined by the 
decomposition of $G$ into mutually commuting subalgebras, $\calG=
\calG_1 \oplus ... \oplus \calG_n$, and is of the form
\beq
\label{scaleip}
\inprod{.}{.} = \frac{1}{q_1^2} \inproda{.}{.}_1 + ... +
\frac{1}{q_n^2} \inproda{.}{.}_n,
\eeq
with $\inproda{.}{.}_i$ the inner product $\inproda{X}{Y} = -p
\tr(XY)$, restricted to $\calG_i$. This has $n$-scales characterising
all possible $G$-invariant inner products on $\calG$. In a gauge
theory context these scales correspond to the gauge coupling constants.

Note that the same symbol is used to denote the inner product on
$\calG$ and $\calV$; we hope it should be clear from the context which
we are using. Corresponding norms for these two inner products are
denoted by $\norm{.}$. We discuss these inner products more fully in
the companion paper \cite{me2}. 

A reference point $\Phi_0 \in \calV$ is arbitrary because of the
degeneracy given by the vacuum manifold $M=D(G)\Ph_0 \cong G/H$. Where
here $H$ is the residual symmetry group, defined by the reference
point $\Ph_0$ to be $H = \{g \in G : D(g) \Ph_0 = \Ph_0\}$. Then
$H$ determines a reductive decomposition of $\calG$  
\beq
\calG = \calH \oplus \m,
\eeq
with $\calH$ the Lie algebra of $H$, such that 
\beq 
[\calH, \calH] \subseteq \calH, \ \ \ \ {\rm and}\ \ \ \ [\calH, \m]
\subseteq \m. 
\eeq
Under the adjoint action of $H$, defined ${\Ad}(h)X =hXh^{-1}$, $\m$
decomposes into irreducible subspaces  
\beq
\m = \m_1 \oplus \cdots \oplus \m_N.
\eeq
These irreducible spaces describe how the group acts on the vacuum
manifold; yielding the family structure for embedded defects. 

Finally, recall that the centre $\calC$ of $\calG$ is the
set of elements that commute with $\calG$. Then the stability of
vortices is related to the projection of $\calC$ onto $\m$,  
\beq
{\rm pr}_\calM(X) = X + X_h,
\eeq
with $X_h \in \calH$ the unique element such that ${\rm pr}_\calM(X)
\in \calM$. One should note ${\rm pr}_\calM(\calC)$ consists of
one-dimensional irreducible $\m_i$'s. 

This structure is enough to categorise all the topological and
non-topological embedded domain wall, embedded vortex and embedded
monopole solutions of a Yang-Mills theory.

\subsection{Domain Walls}

Embedded domain walls are defined elements $\Ph_0 \in \calV$:
\bse
\bea
\Ph(z) &=& f_{\rm DOM}(z) \Ph_0, \\
A^\mu &=& 0, 
\eea
\ese
where $f_{\rm DOM}$ is a real function such that $f_{\rm DOM}(+\infty)
= 1$, and $f_{\rm DOM}(-\infty) \Ph_0 \neq \Ph_0$ belongs to the
vacuum manifold.

Providing the vacuum manifold is connected this solution is unstable;
suffering from a short range instability in the scalar field. Solutions
within connected parts of the vacuum manifold are gauge equivalent.

\subsection{Vortices}

Embedded vortices are defined by pairs $(\Ph_0, X) \in \calV \x
\calM$,
\bse
\label{eq-7}
\bea
\Ph(r,\theta) &=& 
f_{\rm NO}(X;r)D(e^{\theta X})\Ph_0, \\
\underline{A}(r,\theta) &=& 
\frac{g_{\rm NO}(X;r)}{r}X \underline{\hat{\theta}}.
\eea
\ese
Here $f_{\rm NO}$ and $g_{\rm NO}$ are the Nielsen-Olesen profile
functions for the vortex~\cite{Niel78} and we describe their
dependence upon $X$ in the appendix of the companion paper \cite{me2}. 
The vortex generator $X$ has the constraints 
\footnote{there are some complications when the rank (see prenote of
\cite{me2}) of $\m_i$ is greater than one --- we shall generally
indicate when such happens in the text.}
\bse
\bea
X \in \m_i,\\
D(e^{2 \pi X})\Ph_0 = \Ph_0.
\eea
\ese
The winding number of such a  vortex is given by
$\norm{X}/\norm{X^{\min}}$, where $X^{\min}$ is a non-trivial minimal 
generator in the same $\m_i$ as $X$ obeying the above two conditions. 

Family structure originates from the gauge equivalence of vortices 
defined by equal norm generators in the same $\m_i$. 

Vortex stability subdivides into two types: dynamical and topological.
Furthermore, there are two types of topological stability: Abelian,
from $U(1) \rightarrow {\bf 1}$ symmetry breaking; and non-Abelian,
which is otherwise. In \cite{me2} we show that Abelian topological and 
dynamical stability relate to ${\rm pr}_\calM(\calC)$: Abelian
topological stability corresponds to a trivial projection, whilst
dynamical stability corresponds to a non-trivial projection.

Generally, only generators $X \in \calM_i$ define embedded vortices.
However, if the coupling constants $\{q_k\}$ take critical values,
such that between, say, $\m_i$ and $\m_j$,
\beq
\frac{\norm{d(X_i) \Ph_0}}{\norm{X_i}} = 
\frac{\norm{d(X_j) \Ph_0}}{\norm{X_j}}, 
\ \ X_i \in \m_i,\ X_j \in \m_j,
\label{eq-9}
\eeq
then one has extra {\em combination} embedded vortices defined by
generators in $\m_i \oplus \m_j$. 

\subsection{Monopoles}

Embedded monopoles are defined by triplets $(\Ph_0, X_1, X_2) \in
\calV \x \calM \x \calM$:
\bse
\label{eq-10}
\bea
\underline{\Ph}(\underline{r}) &=& f_{\rm mon}(r) \underline{\hat{r}},\\
A^\mu_a(\underline{r}) &=& \frac{g_{\rm mon}(r)}{r}
\ep_{\mu a b} X_b, 
\eea
\ese
where $X_3 = [X_1, X_2]$, and we are treating $\underline{\Phi}$ as a
vector within in its embedded subtheory. 

Monopole generators have the following restrictions~\cite{me2}:\\
(i) The pair $(X_1, X_2) \in \calM_i \x \calM_i$, and are properly
normalised so that, for $i=\{1,2\}$, 
\beq
\exp(2 \pi X_i) \Phi_0 = \Phi_0.
\eeq
(ii) The pair $(X_1, X_2)$ consists of two members of an orthogonal
basis of an $su(2) \subset \calG$, thus
\beq
\norm{X_1} = \norm{X_2}, \ \ \ \ \inprod{X_1}{X_2}=0,
\eeq
and 
\beq
[X_1, [X_1, X_2]] \propto X_2, \ \ \ \ [X_2, [X_1, X_2]] \propto X_1.
\eeq
(iii) the embedded $SU(2)$ is such that $SU(2) \cap H = U(1)$, thus
\beq
[X_1, X_2] \in \calH.
\eeq

The winding number of the monopole is given by
$\norm{X_1}/\norm{X_1^{\min}}$, where $X_1^{\min}$ is the minimal 
generator in the same $\m_i$ as $X_1$ obeying the above conditions. 

Monopoles also have a family structure, depending upon which $\m_i$
they are defined from.


\section{Defects in the Weinberg-Salam Theory}
\label{sec-3}

To illustrate our results we rederive the existence and properties 
of the W and Z-strings~\cite{Vach92-1, Vach92-2} for Weinberg-Salam
theory. One should note that it is the simplest example that
illustrates our formalism. 

The isospin-hypercharge gauge symmetry $G =SU(2)_I \times U(1)_Y$,
acts fundamentally on a two-dimensional complex scalar field $\Ph$. As
a basis we take the $SU(2)$-isospin generators to be
$X^a=\frac{i}{2}\si^a$, with $\si^a$ the Pauli spin matrices,  
and the $U(1)_Y$-hypercharge generator to be $X^0 =\frac{i}{2} {\bf
1}_2$. Then these generators act fundamentally upon the scalar field
$\Phi$ 
\beq
d(\al^iX^i + \al^0 X^0) = \al^i X^i + \al^0 X^0.
\label{eq-39}
\eeq

The inner product on $su(2)_I \oplus u(1)_Y$ may be written
\beq
\label{wsinprod}
\inprod{X}{Y} = -\frac{1}{g^2} \left\{ 2{\tr}XY + (\cot^2 \theta_w-1)
{\tr} X {\tr} Y \right\},
\eeq
with $g$ and $g'$ the isospin and hypercharge gauge coupling
constants. The Weinberg angle $\theta_w = \tan^{-1} (g'/g)$.

Choosing a suitable reference point in the vacuum manifold 
\beq
\Ph_0 = \frac{v}{\sqrt{2}} 
\left( \begin{array}{c} 0 \\ 1 \end{array} \right),
\eeq
the gauge groups breaks to
\beq{
H = U(1)_Q = \left( \begin{array}{cc} e^{i\om} & 0 \\ 0 & 1 \end{array}
\right),} 
\eeq
with $\om \in [0, 2\pi)$. Then $H$ defines the decomposition $\calG =
\calH \oplus \m$, where 
\beq
\calH = \left( \begin{array}{cc} i \al & 0 \\ 0 & 0 \end{array}
\right)\ \ \ \ {\rm and}\ \ \ \ 
\m = \left( \begin{array}{cc} -i\be \cos2 \theta_w & \ga \\ -\ga^* &
i\be \end{array} \right), 
\eeq
with $\al, \be$ real and $\ga$ complex. The star denotes complex
conjugation.  

Under $\Ad(H)$, $\calM$ is reducible to $\m = \m_1 \oplus \m_2$, where
\beq
\m_1 =
\left( \begin{array}{cc} -i\be \cos2 \theta_w & 0 \\ 0 & i \be
\end{array} \right) \ \ \ \ {\rm and}\ \ \ \ 
\m_2 =
\left( \begin{array}{cc} 0 & \ga \\ -\ga^* & 0 \end{array}
\right). 
\eeq
The centre of $su(2)_I \oplus u(1)_Y$, which is $\calC = u(1)_Y$,
projects non-trivially onto $\m_1$ under the inner product
(\ref{wsinprod}). 

The first class of embedded vortices are defined from elements $X \in
\m_1$ such that $e^{2 \pi X} = 1$. Since $\m_1 = {\rm
pr}_\calM(u(1)_Y)$ these vortices are stable in the coupling constant
limit $g \rightarrow 0$. From Eq.~(\ref{eq-7}) one immediately writes
down the solution as:
\bse
\bea
\Ph(r,\theta) &=& \frac{v}{\sqrt{2}} f_{\rm NO}^Z(r) \left(
\begin{array}{c} 0 \\ e^{in\theta} 
\end{array} \right), \\
\underline{A}(r, \theta) &=& \frac{g_{\rm NO}^Z(r)}{r} 
\left( \begin{array}{cc} -in \cos2 \theta_w & 0 \\ 0 & in \end{array}
\right) 
\underline{\hat{\theta}}, 
\eea
\ese
where $n$ is the winding number of the vortex. Note that
this vortex is also invariant under global transformations
of the residual gauge symmetry. These solutions are Z-strings. 

The second class of embedded vortices are defined from elements $X \in
\m_2$ such that $e^{2 \pi X} =  1$. From Eq.~(\ref{eq-7}) one
immediately writes down the solution as:
\bse
\bea
\Ph(r,\theta) &=& \frac{v}{\sqrt{2}} f_{\rm NO}^W(r) \left(
\begin{array}{c} e^{i \de} 
{\sin} n\theta \\ {\cos} n\theta \end{array} \right), \\
d(\underline{A}(r, \theta)) &=& \frac{g^W_{\rm NO}(r)}{r} 
\left( \begin{array}{cc} 0 & ne^{i\de} \\ -ne^{-i\de} & 0 \end{array}
\right) \underline{\hat{\theta}}, 
\eea
\ese
with $e^{i \de} = \ga/{\mid \! \ga \! \mid}$ and $n$ the winding
number of the vortex. All the isolated solutions of the same winding
number  in this one-parameter family are gauge equivalent. 
Furthermore, the anti-vortex is gauge equivalent to the vortex, so
isolated  solutions are parameterised by the positive winding number
only. These solutions are W-strings.

The above generators in $\m_1$ and $\m_2$
satisfy the condition $\norm{d(X) \Phi_0}/\norm{\Phi_0}=n$ of the
Appendix in the companion paper \cite{me2}. Thus, profile
functions for the Z and W-strings are related (first stated in
\cite{MacD95}) 
\bse
\bea
f_{\rm NO}^Z(\la ; r) &=& f_{\rm NO}^W(\frac{\la}{\ka^2} ; \ka r),\\
g_{\rm NO}^Z(\la ; r) &=& g_{\rm NO}^W(\frac{\la}{\ka^2} ; \ka r),
\eea
\ese
where $\ka = \sqrt{\frac{g^2 +g'^2}{g^2}}$ and $\la$ is the quartic
scalar self coupling.

\section{The Model $SU(3) \rightarrow SU(2)$.}
\label{sec-the model su3 -> su2}

We give here an example a model that admits as a solution an unstable 
globally gauge invariant vortex. In addition it is a nice example of a
model admitting non-topological embedded monopoles. 

The gauge group is $G=SU(3)$, acting fundamentally on a
three-dimensional complex scalar field. Denoting the generators by
$\{X^a:a=1 \cdots 8\}$, the derived representation acts as:
\beq
d(\al^i X^i) = \al^i X^i,
\eeq

A Landau potential is sufficient to break the symmetry,
because $\m$ is of the same dimension as the maximal sphere contained
within ${\bf C}^3$. Hence, the vacuum manifold is isomorphic to a
five-sphere, with $G$ transitive over it. 

Taking the reference point in the vacuum manifold to be 
\beq
\Ph_0 = v \left( \begin{array}{c} 0 \\ 0 \\ 1 \end{array} \right),
\eeq
the gauge group breaks to $H=SU(2)$, 
\beq
H = \left( \begin{array}{ccc} SU(2) & \vdots & 0 \\ 
\cdots & \cdots & \cdots \\ 0 & \vdots & 1 \end{array} \right) \subset G.
\eeq
At the reference point $\Ph_0$, $\calG$ decomposes under $\Ad_G(H)$
into irreducible subspaces of the form $\calG = \calH \oplus \m_1
\oplus \m_2$, 
where
\beq
\m_1 = \left( \begin{array}{ccc} i\ga & 0 & 0 
\\ 0 & i\ga & 0 \\ 0 & 0 & -2i\ga \end{array} \right),
\ \ \ {\rm and}\ \ \ 
\m_2 = \left( \begin{array}{ccc} 0 & 0 & a \\ 
0 & 0 & b \\ -a^* & -b^* & 0 \end{array} \right),
\eeq
with $\ga$ real and $a,b$ complex.

The first class of vortex solutions are classified by $X \in \m_1$. 
They are given by
\bse 
\bea
\Ph(r,\theta) &=& v f_{\rm NO}(X_1 ;r) \left( \begin{array}{c} 0 \\  
0 \\ e^{in\theta} \end{array} \right), \\
\underline{A}(r, \theta) &=& \frac{g_{\rm NO}(X_{1} ;r)}{r} 
\left( \begin{array}{ccc} -in/2 & 0 & 0 \\ 0 & -in/2 & 0 \\
0 & 0 & in \end{array} \right)
\underline{\hat{\theta}}. 
\eea
\ese
The integer $n$ is the winding number of the vortex.
These solutions have no semi-local limit and are therefore always
unstable.

The second class of vortex solutions are those classified by $X \in
\m_2$. They are a three-parameter family of gauge equivalent, unstable
solutions. 

The vortex winding number in both classes $\m_1$ and $\m_2$ is
$\norm{d(X) \Phi_0}/\norm{\Phi_0}$. From the Appendix of the companion
paper~\cite{me2}, profile functions for both 
classes coincide with each other and the Abelian-Higgs model.

Non-topological embedded monopole solutions are present in this
model.The solutions are specified by a gauge
equivalent class of generators $(X , Y) \in \m_2 \x \m_2$, such that 
$\inprod{X}{Y} = 0$ and $[X, Y] \in \calH$. A class of such generators
is  
\beq
X = {\Ad}(h) \left( \begin{array}{ccc} 0 & 0 & 1 \\ 0 & 0  & 0 \\
-1 & 0 & 0 \end{array} \right),\ \ \ \ 
Y = {\Ad}(h) \left( \begin{array}{ccc} 0 & 0 & 0 \\ 0 & 0  & 1 \\
0 & -1 & 0 \end{array} \right),
\eeq
with
\beq{
[X, Y] = {\Ad}(h) \left( \begin{array}{ccc} 
0 & -1 & 0 \\ 1 & 0  & 0 \\
0 & 0 & 0 \end{array} \right) \in \calH,}
\eeq
where $h$ is some element in $H$. There is a one-to-one correspondence
between elements in $H$ and the choice of embedded monopole. It should
be noted that elements of the form
\beq
X' = {\Ad}(h) \left( \begin{array}{ccc} 0 & 0 & 1 \\ 0 & 0  & 0 \\
-1 & 0 & 0 \end{array} \right),\ \ \ \ 
Y' = {\Ad}(h) \left( \begin{array}{ccc} 0 & 0 & i \\ 0 & 0  & 0 \\
i & 0 & 0 \end{array} \right),
\eeq 
do not define monopole solutions because $[X', Y'] \not\in
\calH$. Anti-monopoles are defined in the above form but
with one of the generators negative.

In conclusion, there is a two-parameter family of unstable embedded
monopole solutions of the form defined in Eq.~(\ref{eq-10}).


\section{The Model $U(1) \x U(1) \rightarrow 1$.}
\label{the model u1->u1}

This model is presented to illustrate combination vortices. 
By `combination vortices' we mean vortices that are
generated by elements that are not in any of the irreducible spaces
$\m_i$; the vortex generators being instead {\em between}
the spaces.

In section (2), we said that such combination vortices are solutions
providing the coupling constants take a critical set of values. We
illustrate this principle by explicitly finding such solutions in 
the model $U(1) \x U(1) \rightarrow 1$.

The gauge group is $G = U(1)_X \x U(1)_Y$, with elements
\beq
g(\theta, \varphi)
 = \left( \begin{array}{cc} e^{i\theta} & 0 \\ 0 & e^{i\varphi} \end{array}
\right) \in G, 
\eeq
and $\theta, \varphi \in [0, 2\pi)$.
Generators of $U(1)_X$ and $U(1)_Y$ are
\beq{
X = \left( \begin{array}{cc} i & 0 \\ 0 & 0 \end{array}
\right),\ \ \ \ 
Y = \left( \begin{array}{cc} 0 & 0 \\ 0 & i \end{array}
\right).}
\eeq
The group $G$ acts fundamentally on a two-dimensional complex
scalar field $\Ph = (\ph_1, \ph_2)^\top$
\beq
D(g(\theta, \varphi)) = 
\left( \begin{array}{cc} e^{i\theta} & 0 \\ 0 & e^{i\varphi}
\end{array} \right).
\eeq

The inner product on $\calG$ is of the form
\beq
\inprod{X}{Y} = -\tr(XQ^{-1}Y),\ \ \ {\rm with} \ \ \ 
Q = \left( \begin{array}{cc} q_1^2 & 0 \\ 0 & q_2^2
\end{array}\right), 
\eeq
where $q_1$ and $q_2$ are the coupling constants for the respective
parts of $G$. 

To break $G$ to triviality, the parameters of the scalar
potential must be chosen correctly. The general, renormalisable,
gauge invariant scalar potential for this theory is
\beq
V(\ph_1, \ph_2) = \la_1(\ph_1^* \ph_1 - v_1^2)^2 + \la_2(\ph_2^* \ph_2
- v_2^2)^2 + \la_3 \ph_1^* \ph_1 \ph_2^* \ph_2.
\eeq
For some range of $(\la_1, \la_2, \la_3, v_1, v_2)$ (the range being
unimportant to our arguments) this is minimised by a two-torus 
of values, then $G$ breaks to triviality.

Without loss of generality the scalar field reference point is chosen
to be  
\beq
\Ph_0 = \left( \begin{array}{c} v'_1 \\ v'_2 \end{array} \right),
\eeq
where unless $v_1^2 = v_2^2$, the primed vevs $v_1'$, $v_2'$ are
unequal to $v_1$ and $v_2$.
Then the group $G$ breaks to the trivial group $H = {\bf 1}$. Under
the adjoint action of $H$, the Lie algebra of $G$ splits into
\beq{
\calG = \m_1 \oplus \m_2,}
\eeq
with
\beq{
\m_1 = \left( \begin{array}{cc} ia & 0 \\ 0 & 0 \end{array}
\right),\ \ \ \
\m_2 = \left( \begin{array}{cc} 0 & 0 \\ 0 & i b \end{array}
\right),}
\eeq
and $a, b$ real.

The topology of the vacuum manifold is non-trivial, hence
vortex solutions that are generated by
elements in $\m_1$ or $\m_2$ are topologically stable. These vortices
are well defined and are stationary solutions of the Lagrangian.

It is interesting to consider the existence of vortices
generated by elements in the whole of 
$\m_1 \oplus \m_2$, and not just vortices
generated in either of these two spaces separately. Combination
vortices may exist when the coupling constants are such that
Eq.~(\ref{eq-9}) is satisfied. Substitution of the generators $X$ and
$Y$ into Eq.~(\ref{eq-9}) yields the condition that combination
vortices exist for
\beq
\frac{\norm{d(X)}}{\norm{X}} = \frac{\norm{d(Y)}}{\norm{Y}} 
\Rightarrow 
q_1^2 = q_2^2.
\eeq

When $q_1 = q_2$, the Lie algebra elements that generate closed
geodesics are of the form 
\beq{
Z = \de X + \ep Y,}
\eeq
providing there exists $\om >0$ with $D(e^{Z \om}) \Ph_0 = \Ph_0$.
Since the coupling constants are equal $Z$ generates a
$U(1)$-subgroup of $G$. Relating this back to the
geometry of a torus the constraint on non-zero $\ep$ and $\de$ is
\beq
\frac{\ep}{\de} \in {\bf Q},
\eeq
the rational numbers. One can interpret the effect of the scaling as
`twisting' directions in the tangent space to the vacuum manifold
relative to directions in the Lie algebra. This twisting only happens
between the irreducible subspaces of $\m$. 

However, not all of these geodesics define embedded vortices. One also
needs to satisfy cond. (2) in the companion paper \cite{me2},
\beq
\langle \Psi, \pderiv{V}{\Phi} \rangle =0,
\eeq
where $\Psi \in \calV_\emb^\perp$ and $\Phi \in \calV_\emb$. Trivial
substitution yields
\beq
\la_1 = \la_2 = \la, \ v_1^2 = v_2^2 = v^2, \ {\rm and} \ \ep = \de.
\eeq
This is the only combination vortex.


\section{Embedded Defects in Realistic GUT models}
\label{sec7}

We now gives some examples of the embedded defect spectrum in some
realistic GUT models. The examples here are certainly not meant to be
exhaustive, merely just a few of the simplest examples.

\subsection{Georgi-Glashow $SU(5)$}

The gauge group is $G= SU(5)$ \cite{Geor74}, acting on a twenty-four
dimensional scalar field $\Ph$ by the adjoint action. For scalar vacuum,
\beq
\Ph_0 = v \left( \begin{array}{ccc} \frac{2}{3} \bf{1}_3 & \vdots &
\bf{0} \\ \cdots & \cdots  & \cdots \\ \bf{0} & \vdots & -\bf{1}_2
\end{array} \right), 
\eeq
$G$ breaks to $H = SU(3)_C \x SU(2)_I \x U(1)_Y$,
\beq
\left( \begin{array}{ccc} SU(3)_0 & \vdots &
\bf{0} \\ \cdots & \cdots  & \cdots \\ \bf{0} & \vdots & SU(2)_I
\end{array} \right)
\x
\left( \begin{array}{ccc} e^{\frac{2}{3}i \theta} {\bf 1_3} & \vdots &
\bf{0} \\ \cdots & \cdots  & \cdots \\ \bf{0} & \vdots & 
e^{-i \theta}{\bf 1}_2 \end{array} \right)
\subset SU(5).
\eeq

To find the embedded defect spectrum one determines the
reduction of $\calG$ into $\calG = \calH \oplus \m$ and finds the
irreducible spaces of $\m$ under the adjoint action of $H$. The space
$\m$ is 
\beq
\m = \left( \begin{array}{ccc} \bf{0}_3 & \vdots &
\underline{A} \\ \cdots & \cdots  & \cdots \\ -\underline{A}^\dagger 
 & \vdots & \bf{0}_2
\end{array} \right),
\eeq
with $\underline{A}$ a two-by-three complex matrix. This is
irreducible under the adjoint action of $H$. 

Thus the defect spectrum of the model is: monopoles, which can be
confirmed to be topologically stable; and a family
of unstable Lepto-quark strings. The family of lepto-quark strings is
complicated by $\m$ containing {\em two} distinct (non-proportional)
commuting generators.

\subsection{Flipped-$SU(5)$}
\label{flip}

For a more detailed discussion of embedded defects and their
properties in flipped-$SU(5)$, see \cite{me1}.

The gauge group is $G=SU(5) \x \widetilde{U(1)}$ \cite{Barr82}, and
acts upon a complex ten dimensional scalar field (which we conveniently
represent as a five by five, complex antisymmetric matrix) by the {\bf
  10}-antisymmetric representation. Denoting the generators of $SU(5)$
as $X^a$ and $\widetilde{U(1)}$ as $\widetilde{X}$, the derived 
representation acts upon the scalar field as
\beq
d(\al^i X^i + \al^0 \widetilde{X}) 
= \al^i (X^i \Ph + \Ph {X^i}^\top) + \al^0 \widetilde{X}
\Ph.
\eeq

The inner product upon $su(5) \oplus u(1)$ is of the form:
\beq
\label{inprodflpd}
\inprod{X}{Y} = -\frac{1}{g^2} \left\{ {\tr}XY + \frac{1}{5}(\cot^2
\Theta-1) {\tr} X {\tr} Y \right\},
\eeq
where $g$ and $\tilde{g}$ are the $SU(5)$ and $\widetilde{U(1)}$
coupling constants. The GUT mixing angle is $\tan \Theta =
\tilde{g}/g$ 

For the following discussion it is necessary to explicitly know the 
following generators 
\beq
X^{15} = ig \sqrt{\frac{3}{2}} 
\left( \begin{array}{ccc}  \frac{2}{3} \bf{1}_3 & \vdots & \bf{0} \\ 
\cdots & \cdots & \cdots \\
\bf{0} & \vdots & - \bf{1}_2
\end{array} \right),\ \ \ \ 
\widetilde{X} = i \tilde{g} \bf{1}_5.
\eeq
These generators are normalised with respect to~(\ref{inprodflpd}).

For a vacuum given by
\beq
\Phi_0 = {v \over \sqrt{2}} 
\left( \begin{array}{ccc} \bf{0}_3 & \vdots & \bf{0} \\ 
\cdots & \cdots & \cdots \\
\bf{0} & \vdots & I
\end{array} \right),\ \ \ {\rm where}\ \ \ 
I = 
\left( \begin{array}{cc} 0 & 1 \\ 
-1 & 0
\end{array} \right),
\eeq
one breaks $SU(5) \x U(1)$ to the standard model $H = SU(3)_C \x
SU(2)_I \x U(1)_Y$, provided that the parameters of the potential
satisfy $\et^2, \la_1 >0$ and $(2 \la_1 + \la_2) > 0$.
The V and hypercharge fields are given by 
\bse
\bea
V_\mu = \cos \Theta A_\mu^{15} - \sin \Theta \widetilde{A}_\mu, & &
X_V = \cos \Theta X^{15} - \sin \Theta \widetilde{X},\\ 
Y_\mu = \sin \Theta A_\mu^{15} + \cos \Theta \widetilde{A}_\mu, & &
X_Y = \sin \Theta X^{15} + \cos \Theta \widetilde{X}.
\eea
\ese
Then $d(\calH) \Ph_{\rm vac} = 0$. The isospin and colour symmetry
groups are 
\beq
\left( \begin{array}{ccc}  SU(3)_C & \vdots & \bf{0} \\ 
\cdots & \cdots & \cdots \\
\bf{0} & \vdots & SU(2)_I
\end{array} \right) \subset SU(5).
\eeq

To find the embedded defect spectrum one determines the
reduction of $\calG$ into $\calG = \calH \oplus \m$ and finds the
irreducible spaces of $\m$ under $\Ad(H)$: which is $\calM = \calM_1
\oplus \calM_2$ such that 
\beq
\m_1 = {\bf R} X_V,\ \ \ \ \ \ 
\m_2 = 
\left( \begin{array}{ccc}  \bf{0}_3 & \vdots & \underline{A} \\ 
\cdots & \cdots & \cdots \\
-\underline{A}^\dagger & \vdots & \bf{0}_2
\end{array} \right).
\eeq

The first space $\m_1$ is the projection of
$\widetilde{u(1)}$ onto $\m$. This is important for the
stability of vortex solutions defined from it. Such vortices
are stable in the limit $\Theta_{\rm GUT} \rightarrow \frac{\pi}{2}$,
then, by continuity, also in a region around $\frac{\pi}{2}$.

The second space $\m_2$ generates a family of unstable Lepto-quark
strings and non-topological monopoles. The family of lepto-quark
strings is complicated by $\m$ containing {\em two} distinct
(non-proportional) commuting generators.

\subsection{Pati-Salam $SU(4) \x SU(4) \rightarrow SU(3)_0 \x SU(2)_I
  \x U(1)_Y$}

Pati and Salam emphasised a series of models of the form $G=G^S \x G^W$,
where $G^S$ and $G^W$ are identical strong and weak groups related by
some discrete symmetry~\cite{Pati73}. The above model is the 
simplest one of this form. The model is actually $[SU(4) \x SU(4)]_L
\x [SU(4) \x SU(4)]_R$ (`L' and `R' denoting the separate couplings to
left and right-handed fermions) to accommodate parity violation in weak
interactions. For simplicity we shall only consider half of the model.

The gauge group $G= SU(4)^S \x SU(4)^W$, breaking to $H=(SU(3)
\x U(1))^S \x SU(2)^W$, 
\beq
\left( \begin{array}{ccc}  SU(3)_C & \vdots & \bf{0} \\ 
\cdots & \cdots & \cdots \\
\bf{0} & \vdots & U(1)_Y
\end{array} \right)_S \x
\left( \begin{array}{ccc}  SU(2)_I & \vdots & \bf{0} \\ 
\cdots & \cdots & \cdots \\
\bf{0} & \vdots & \bf{1}_2
\end{array} \right)_W \subset SU(4)^S \x SU(4)^W.
\eeq
Writing $\calG = \calH \oplus \m$, the irreducible
spaces of $\m$ under $\Ad(H)$ are $\m =
\m_1 \oplus \m_2 \oplus \widetilde{\m}$, where, $\widetilde{\m}$ is a
collection of four irreducible spaces, with
\bea
\m_1 = \left( \begin{array}{ccc} \bf{0}_3 & \vdots & \underline{A} \\ 
\cdots & \cdots & \cdots \\
-\underline{A}^\dagger & \vdots & 0 \end{array} \right)_S,\ \ \ \ \ \ 
\m_2 = \left( \begin{array}{ccc} \bf{0}_2 & \vdots & B \\ 
\cdots & \cdots & \cdots \\
-B^\dagger & \vdots & \bf{0}_2 \end{array} \right)_W,\nn \\
{\rm and}\ \ \ \ \ \ 
\widetilde{\m} = \left( \begin{array}{ccc} \bf{0}_2 & \vdots &
\bf{0}_2 \\ \cdots & \cdots & \cdots \\ 
\bf{0}_2 & \vdots & C \end{array} \right)_W \oplus
\left( \begin{array}{ccc} i \al \bf{1}_2 & \vdots &
\bf{0}_2 \\ \cdots & \cdots & \cdots \\ 
\bf{0}_2 & \vdots & -i \al \bf{1}_2 \end{array} \right)_W,
\eea
where $\underline{A}$ is a complex three dimensional vector, $B$
and $C$ are complex two by two matrices, with $C$ anti-hermitian, and
$\al$ is a real number. 

Each of the above spaces gives rise to their respective embedded
defects. Firstly, $\m_1$ gives rise to topologically stable monopoles
and a five parameter family of unstable vortices. Secondly, 
$\m_2$ gives rise to non-topological
unstable monopoles and an seven parameter family of
unstable vortices. Thirdly, $\widetilde{\m}$, which is a collection of
four irreducible spaces, admits globally gauge invariant unstable
vortices. In addition, $\widetilde{\m}$ has combination vortex
solutions between the four irreducible spaces that it consists of.


\section{Vortices in the $^3$He-A Phase Transition}
\label{sec8}

We wish to show here that our results on the classification of
vortices for general gauge theories are also relevant for
condensed matter systems. As an example we choose the $^3$He-A
phase transition, though we expect the general onus of our results to
be applicable to other situations having a similar nature.

Superfluid $^3$He has global symmetries of spin ($SO(3)_S$ rotations),
angular 
rotations ($SO(3)_L$) and a phase (associated with particle number
conservation). It has several phase transitions corresponding to
different patterns of breaking this symmetry. We concentrate here on
the A-phase transition. 

Condensed matter systems, such as $^3$He, have added complications
above that of gauge theories, meaning that we cannot just naively
apply the 
approach used in the rest of this paper. This complication originates
through the order parameter being a {\em vector} under spatial
rotations, not a scalar as in conventional gauge theories. The upshot
being that extra terms are admitted in the Lagrangian that are not
present in a conventional gauge theory. These terms couple
derivatives of components with different angular momentum quantum
numbers and are so not invariant under $SO(3)_L$ rotations in the
conventional sense --- thus spoiling the $SO(3)_L$ invariance.
The general effect of this is to complicate the spectrum of
vortex solutions, and their actual form and
interaction.

Our tactic to investigate the effect of these extra non-invariant
$SO(3)_L$ terms is to firstly examine the $^3$He-A phase transitions
without inclusion of these terms so that we may use the techniques of
embedded vortices used in the rest of this paper, and then to see how
these terms affect the solutions.

\subsection{The $^3$He-A Phase Transition}

The full symmetry group of liquid $^3$He is
\beq
G_{3He} = SO(3)_S \x SO(3)_L \x U(1)_N,
\eeq
which acts on the two group-index order parameter $A_{\al j}$ by the
{\em fund.}$_S \otimes$ {\em fund.}$_{L,N}$ representation of
$G_{3He}$. Denoting
\beq
A_{\al j} = \De_0 d_\al \Psi_j,
\eeq
with unit vector $d_\al \in {\bf R}^3$ and $\Psi_j = (\hat{\bf e}_1
+i\hat{\bf e}_2)/\sqrt{2} \in {\bf C}^3$, where $\hat{\bf
e}_1,\hat{\bf e}_2 \in {\bf R}^3$ such that $\hat{\bf e}_1 . \hat{\bf
e}_1 = \hat{\bf e}_2 . \hat{\bf e}_2 = 1$ and $\hat{\bf e}_1.\hat{\bf
e}_2=0$. The quantity $\De_0$ is a real number unimportant for the
present discussion. 

Then $G_{3He}$ acts on $A_{\al j}$ fundamentally:
\beq
D( (g_S, g_L, g_N))_{\al j \be k} A_{\be k} = \De_0 (g_S {\bf d})_\al
(g_L g_N {\bf \Psi})_j.
\eeq
In addition $G_{3He}$ is a global symmetry of the field theory.

The field theory is described by the Lagrangian
\beq
\calL[A_{\al j}] = \calL_{\rm sym}[A_{\rm \al j}] +
\widetilde{\calL}[A_{\al j}], 
\eeq
with $\calL_{\rm sym}$ having $G_{3He}$ global symmetry and
$\widetilde{\calL}$ 
representing the extra vector type couplings of the order
parameter. We may write
\beq
\calL_{\rm sym}[A_{\rm \al j}] = \ga \partial_i A_{\al j}^\star
\partial_i A_{\al j} - V[A_{\al j}],
\eeq
with $V$ some Landau-type potential invariant under $G_{3He}$. The
vector-type couplings we write
\beq
\widetilde{\calL}[A_{\al j}] = \ga_1 \partial_i A_{\al i}^\star
\partial_j A_{\al j} + \ga_2 \partial_i A_{\al j}^\star \partial_j
A_{\al i}, 
\eeq
which are explicitly not $SO(3)_L$ invariant. By partial integration
of the {\em action} integral, this may be rewritten as
\beq
\widetilde{\calL}[A_{\al j}] = (\ga_1 + \ga_2) \partial_i A_{\al i}^\star
\partial_j A_{\al j} =  \tilde{\ga} \partial_i A_{\al i}^\star
\partial_j A_{\al j}.
\eeq

The A-phase is reached through symmetry breaking with a vacuum of the
form
\beq
A_0 = \De_0 {\bf d}_0 {\bf \Psi}_0, \ {\rm where}\ 
{\bf d}_0 = \left( \begin{array}{c} 1 \\ 0 \\ 0 \end{array}
\right),\ 
{\bf \Psi}_0 = \left( \begin{array}{c} 1 \\ i \\ 0 \end{array}
\right),
\eeq
so that the residual symmetry group is
\beq
H_A = U(1)_{S_3} \x U(1)_{L-N}  \x {\bf Z}_2,
\eeq
where
\bse
\bea
U(1)_{S_3} &=& \left\{ 
\left( \begin{array}{ccc} 1 & 0 & 0 \\ 0 & \cos \al & \sin \al \\
0 & - \sin \al& \cos \al \end{array} \right)_S
: \ \al \in [0, 2\pi) \right\} \\
U(1)_{L-N} &=& \left\{ e^{-i \al}
\left( \begin{array}{ccc} \cos \al & \sin \al & 0 \\
-\sin \al & \cos \al & 0 \\ 0 & 0 & 1 \end{array} \right)_L
: \ \al \in [0, 2\pi) \right\} \\
{\bf Z}_2 &=& \left\{ {\bf 1}_S \x {\bf 1}_L,\ 
\left( \begin{array}{ccc} -1 & 0 & 0 \\ 0 & -1 & 0 \\
0 & 0 & 1 \end{array} \right)_S \x
\left( \begin{array}{ccc} -1 & 0 & 0 \\ 0 & -1 & 0 \\
0 & 0 & 1 \end{array} \right)_L \right\}
\eea
\ese
It should be noted that the $\{L,N\}$ part of the
group is similar to the Weinberg-Salam theory at $\Theta_w = \pi/4$, but
taking the limit in which (both) of the coupling constants become
zero. However, note that $SO(3)_L$ is {\em not} simply connected;
this has important stabilising effects on the vortices
\cite{Volo95}.  

Writing $\calG_{3He} = \calH_A \oplus \m$, the irreducible spaces of $\m$
under the adjoint action of $H_A$ are denoted by $\m =
\m_1 \oplus \m_2 \oplus \m_3$, with 
\bea
\m_1 = \left( \begin{array}{ccc} 0 & 0 & \al \\ 
0 & 0 & \be \\ -\al & -\be & 0 \end{array} \right)_L,\ \ \ \ \ \ 
\m_2 = \left( \begin{array}{ccc} 0 & \ga & \de \\ 
-\ga & 0 & 0 \\ -\de & 0 & 0 \end{array} \right)_S,\nn \\
{\rm and}\ \ \ \ \ \ 
\m_3 = \frac{\ep}{2} \left( \begin{array}{ccc} i & 1 & 0 \\ 
-1 & i & 0 \\ 0 & 0 & i \end{array} \right)_L, 
\eea
and $\al, \be, \ga, \de, \ep$ are real numbers.

\subsection{Vortices in the $SO(3)_L$ Symmetric Theory}

We firstly analyse the theory when $\tilde{\ga} = 0$, so that the
Lagrangian is $SO(3)_L$ symmetric. In this regime the techniques of
embedded vortices are applicable.

\subsubsection{Embedded Vortices}

The first class of generators, $\m_1$, give a one parameter family of
gauge equivalent global vortices, with profiles of the form
\beq
A(r, \theta ) = \De_0 \bar{f}(n/\sqrt{2}; r) {\bf d}_0
\left( \begin{array}{c} \cos \al/2 + i \sin \al/2 \cos n \theta \\ -\sin
  \al/2 + i \cos \al/2 \cos n \theta \\ -i \sin n \theta \end{array}
\right).
\eeq
Here $n$ is the winding of the vortex, $\al$ labels the family
member, and $\bar{f}$ is defined below. These are the disgyration
vortices of $^3$He. 

The second class of generators, $\m_2$, give a one parameter family of
gauge equivalent global vortices, with profiles of the form
\beq
A(r, \theta) = \De_0 \bar{f}(n;r)
\left( \begin{array}{c} \cos n \theta \\ - \cos \al \sin n \theta \\ 
  \sin \al \sin n \theta \end{array}
\right) {\bf \Psi}_0.
\eeq
Here $n$ is the winding of the vortex, and $\al$ labels the family
member. These are the, so called, spin vortices.

The third class of generators, $\m_3$, give a 
gauge invariant global vortex, with a profile of the form
\beq
A(r, \theta) = \De_0 \bar{f}(n;r) {\bf d}_0 e^{i n \theta}
\left( \begin{array}{c} 1 \\ i \\ 0 \end{array} \right).
\eeq
Here $n$ is the winding of the vortex, and $\al$ labels the family
member. These vortices are the, so called, singular-line vortices.

The profile functions depend upon the embedded vortex considered,
generated by $X_\emb$ say, and are minima of the Lagrangian 
\beq
\calL[f] = \frac{\ga \De_0^2}{2} \left( \deriv{f}{r} \right)^2 +
\frac{\ga f^2}{2 r^2} \norm{X_\emb A_0}^2 -
V[f(r)],  
\eeq
where $V$ is the potential, which is independent of the defect
considered. Writing $\norm{X_\emb A_0}=n\norm{A_0}$ we refer to the
solutions as $\bar{f}(n;r)$.

\subsubsection{Combination Vortices}

Because the symmetries $G_{3He}$ are global there are combination
vortex solutions between the three families of generators. The most
general combination embedded vortex is generated by a combination of
generators from each of the three classes --- this is the spin -
singular line - disgyration combination vortex. Because of the way we
shall determine such vortices we firstly discuss the singular line
-disgyration combination.

One obtains a discrete spectrum of singular line-disgyration
combination embedded vortices. Solutions are of the form
\bse
\bea
A(r, \theta) &=& \De_0 f(X;r) {\bf d}_0 \exp(X \theta) {\bf
  \Psi}_0, \\
{\rm with \ } X &=& \frac{a}{2} \left( i {\bf 1}_3 +
\left( \begin{array}{ccc} 0 & 1 & 0 \\ 
-1 & 0 & 0 \\ 0 & 0 & 0 \end{array} \right)_L \right) +
b \left( \begin{array}{ccc} 0 & 0 & 1 \\ 
-0 & 0 & 0 \\ -1 & 0 & 0 \end{array} \right)_L.
\eea
\ese
Then some algebra yields
\beq
A(r, \theta) = \De_0 \bar{f}(p;r) {\bf d}_0 e^{ia \theta /2}
\left( \begin{array}{c} \cos \theta s + \frac{ia}{2s} \sin \theta s \\
-\frac{a}{2s} \sin \theta s + \frac{i}{s^2}(b^2 + \frac{a^2}{4} \cos
\theta s) \\
-\frac{b}{s} \sin \theta s + \frac{iab}{2s^2}(\cos \theta s -1)
\end{array} \right),
\eeq
where $s = \sqrt{a^2/4 + b^2}$ and $p= \sqrt{(7m^2+n^2)/2}$. Using the
single valuedness constraint
that $A(r, 2\pi ) = A(r,0)$ gives the following discrete spectrum of
values for $a$ and $b$:
\beq
a= 2m,\ \ \ \ b=\pm \sqrt{n^2 - m^2},\ \ \ \ m,n \in {\bf Z}.
\eeq

It seems that the singular line vortex and the
disgyration may not be continuously deformed into one another, since
if this was 
to be the case then the spectrum of combination vortices should be
{\em continuous}. We obtain a {\em discrete} spectrum. For
them to be continuously deformable into one another
we need solutions that are {\em not} of the embedded type. 

The spin - singular line - disgyration combination vortex can be
constructed from the above form. Since the generators for spin
vortices commute with the generators for singular line - disgyration
combination vortices, the form of solution is a spin vortex combined
with a singular line - disgyration combination, {\em i.e.}
\bea
A(r, \theta) = &\De_0& \bar{f}(\sqrt{(7m^2+2j^2+n^2)/2};r) 
\left( \begin{array}{c} \cos j\theta \\ 
- \cos \al \sin j\theta \\ -\sin \al \sin j\theta \end{array} \right)
\nn \\
& & e^{ia \theta /2} 
\left( \begin{array}{c} \cos \theta s + \frac{ia}{2s} \sin \theta s \\
-\frac{a}{2s} \sin \theta s + \frac{i}{s^2}(b^2 + \frac{a^2}{4} \cos
\theta s) \\
-\frac{b}{s} \sin \theta s + \frac{iab}{2s^2}(\cos \theta s -1)
\end{array} \right),
\eea
with $a$ and $b$ as above and $j$ an integer. Again the spectrum is
discrete. 

In particular, we shall need to know the form of the spin -
singular line  combination embedded vortex, which is:
\beq
A(r, \theta) = \De_0 f(\sqrt{(j^2+n^2)};r) 
\left( \begin{array}{c} \cos j\theta \\ 
- \cos \al \sin j\theta \\ -\sin \al \sin j\theta \end{array} \right)
e^{i n \theta}
\left( \begin{array}{c} 1 \\ i \\ 0 \end{array} \right).
\eeq

\subsubsection{Stability of the Embedded Vortices}

The topology of the vacuum manifold contains loops which are
incontractible and thus gives classes of
stable vortices. With each of the families of embedded (and
combination) vortices an element of the homotopy group may be
associated 
\footnote{more precisely, with the family {\em and} the winding
number, but we shall only be considering unit winding number
vortices}
which tells one whether that family of vortices is topologically
stable or unstable.

The vacuum manifold looks like
\beq
 \frac{SO(3)_S \x SO(3)_L \x U(1)_N}{U(1)_{S_3} \x
U(1)_{L-N} \x {\bf Z}_2}  =
 \frac{S^{(2)}_S \x S^{(3)}_{L,N}/{\bf Z}_2}{{\bf Z}_2}.
\eeq
Here $S^{(n)}$ is an n-sphere. This vacuum manifold contains {\em
three} inequivalent families of incontractible loops. Firstly, those
contained within just $S^{(3)}_{L,N}/{\bf Z}_2$. Secondly, those going
from the identity, through $S^{(2)}_S$ into $S^{(3)}_{L,N}/{\bf Z}_2$
by the ${\bf Z}_2$ factor, and then back to the identity. Thirdly,
there are combination of the first two types. The classes of the first
homotopy group of the vacuum manifold are thus
\beq
\pi_1 \left(\frac{SO(3)_S \x SO(3)_L \x U(1)_N}{U(1)_{S_3} \x
U(1)_{L-N} \x {\bf Z}_2} \right)= {\bf Z}_4.
\eeq

This gives rise to three different
topological charges for the vortices, the charge labelling the
family from which they originate.
Technically, the ${\bf Z}_4$ arises from two separate ${\bf Z}_2$
contributions, and then we can label the charge $(p,q)$, with $p,q =
0,1$; however, a more convenient notation (which will be better
contextualised in the conclusions) is to assign a single index to
these as in \cite{Volo95}, $\nu$: $(0,0) =0, (1,0)=1/2, (0,1)=1,
(1,1)=3/2=-1/2$. 

The $\nu = 1/2$ stable vortices are half-quantum spin - (singular line
- disgyration) combinations --- where one makes use of the ${\bf Z}_2$
mixing of the spin and angular groups for stability. Considering the
spin - singular line combination above (Eq.~(74)), the stable
half-quantum spin-singular line  combination vortex corresponds to
$j=n= \frac{1}{2}$: 
\beq
A(r, \theta) = \De_0 \bar{f}(1/\sqrt{2};r) 
\left( \begin{array}{c} \cos \theta/2 \\ 
- \cos \al \sin \theta/2 \\ -\sin \al \sin \theta/2 \end{array}
\right)
e^{i \theta/2}
\left( \begin{array}{c} 1 \\ i \\ 0 \end{array} \right),
\eeq
Of course, there are also half-quantum spin - disgyration vortices,
and combinations in between. These all have topological charge
$\nu = 1/2$.

The $\nu = 1$ stable vortices are some of the singular line
(Eq.~(68)) 
and disgyration embedded vortices (Eq.~(66)), also including
the combination vortices (Eq.~(71))
inbetween. These all have the form above. 
The winding number $n=1$ vortices are the only stable solutions. 
Odd-$n$ vortices may decay to these, also having topological charge
$\nu=1$; even-$n$ decay to the vacuum, having topological charge
$\nu=0$.

Finally, the $\nu=3/2$ vortices are combinations of the $\nu=1/2$ and
$\nu=1$ vortices. 

\subsection{Vortex Spectra of the Full $^3$He Theory}

We wish to find the embedded vortex spectrum of the 
full $^3$He theory, when
one is including terms which are not invariant under spatial
rotations of the Lagrangian.
Our tactic is to see which of the above embedded vortex solutions
remain solutions in the full theory. This is
facilitated by investigating how the profile equations
 are modified by inclusion of terms that are
not invariant under $SO(3)_L$ --- if the profile equations make sense,
for instance they must only be radially dependent, then one can say
that those embedded vortices remain solutions to the theory.

Providently, it transpires that only those embedded vortices which
are {\em topologically stable} remain solutions to the full $^3$He
Lagrangian with inclusion of terms that are not rotationally
symmetric. 

\subsubsection{Singular-Line Vortices}

The singular-line vortex has a profile of the form (from
Eq.~(68)) 
\beq
A(r, \theta) = \De_0 f(n;r) {\bf d}_0 e^{i n \theta}
\left( \begin{array}{c} 1 \\ i \\ 0 \end{array} \right),
\eeq
where $n$ is the winding number of the vortex. Substitution into the
full Lagrangian (Eq.~(58)) yields the profile equation to be
\beq
\calL[f] + \widetilde{\calL}[f] =
(2\ga + \tilde{\ga}) \De_0^2 \left( \left(\deriv{f}{r} \right)^2 +
\frac{n^2 f^2}{r^2} \right) - 2\tilde{\ga} \De_0^2 \frac{n
f}{r} \deriv{f}{r} - V[f(r)].
\eeq
Since the extra term $nff'/r$ is least dominant asymptotically we may
conclude the the singular line {\em Ansatz} is still a solution to the
full Lagrangian, but with a slightly modified profile
function.

\subsubsection{Spin Vortices}

Vortices embedded solely in the spin sector (with profiles given by
Eq.~(67)) are solutions to the full
Lagrangian because the embedded defect formalism is applicable to
symmetry-invariant parts of the Lagrangian --- which the spin sector
is. 

This observation is backed up within the mathematics; one may show
that for the spin vortex {\em Ansatz}
\beq
\partial_i A_{\al i}^\star \partial_j A_{\al j} =
\partial_i A_{\al j}^\star \partial_i A_{\al j}.
\eeq 
Thus the terms of $\widetilde{\calL}$ that are not invariant under
spatial rotations become equivalent to the kinetic terms of the
symmetric $^3$He Lagrangian for spin vortices.

\subsubsection{Disgyration Vortices}

The embedded disgyration vortex has a profile of the form in Eq.~(66);
to simplify the matter we shall consider the family member with $\al =
0$ (without loss of generality)
\beq
A(r, \theta ) = \De_0 f(n; r) {\bf d}_0
\left( \begin{array}{c} 1 \\ i\cos n \theta \\ -i \sin n \theta
\end{array} \right).
\eeq
where $n$ is the winding of the vortex. Substitution into the full
Lagrangian (Eq.~(58)) yields terms that are not invariant under spatial
rotations
\beq
\widetilde{\calL}[f] = \tilde{\ga} \De_0^2 \left( \left( \cos \theta
\deriv{f}{r}\right)^2 + \left(\cos n \theta \sin \theta \deriv{f}{r}
- \frac{n f}{r} \cos \theta \sin n \theta \right)^2 \right).
\eeq
Since the profile function $f(r)$ is independent of
$\theta$, and the Lagrangian $\calL_{\rm sym}[f] +
\widetilde{\calL}[f]$ that describes $f(r)$ is not rotationally
symmetric, we conclude that the embedded disgyration vortices do not
remain a solution when non-spatially rotationally symmetric terms are
added to the Lagrangian.

\subsubsection{Combination Vortices}

In general only combinations of embedded vortices that individually
remain solutions when non-spatially symmetric terms are added to the
Lagrangian remain solutions. Thus the only combination embedded
vortices that are solutions to the full Lagrangian $\calL_{\rm sym} +
\widetilde{\calL}$ are the {\em combination spin-singular line
vortices}.

\subsection{Conclusions}

We conclude, by comparing the results of sec. (7.3.3) with
sec. (7.2.3), that embedded vortices that are solutions when terms 
rotationally non-symmetric terms are added to the Lagrangian,
\beq
\widetilde{\calL}[A_{\al j}] = (\ga_1 + \ga_2) \partial_i A_{\al i}^\star
\partial_j A_{\al j} =  \tilde{\ga} \partial_i A_{\al i}^\star
\partial_j A_{\al j}, 
\eeq
are those vortices that are topologically stable, or higher winding
number counterparts of those vortices. The topologically stable
embedded vortices are labelled by their topological charge $\nu$
\cite{Volo95} and take the following forms.

Firstly, the half-quantum spin-singular line combination vortex, which
has topological charge $\nu=1/2$ and looks like
\beq
A(r, \theta) = \De_0 \bar{f}(1/\sqrt{2};r) 
\left( \begin{array}{c} \cos \theta/2 \\ 
- \cos \al \sin \theta/2 \\ -\sin \al \sin \theta/2 \end{array}
\right)
e^{i \theta/2}
\left( \begin{array}{c} 1 \\ i \\ 0 \end{array} \right).
\eeq

Secondly, the singular line vortex, which has topological charge
$\nu=1$ and looks like
\beq
A(r, \theta) = \De_0 \bar{f}(1;r) {\bf d}_0 e^{i n \theta}
\left( \begin{array}{c} 1 \\ i \\ 0 \end{array} \right).
\eeq

Thirdly and finally, the combination of the above two vortices, which
has topological charge $\nu=3/2$ and looks like
\beq
A(r, \theta) = \De_0 \bar{f}(\sqrt{5/2};r) 
\left( \begin{array}{c} \cos \theta/2 \\ 
- \cos \al \sin \theta/2 \\ -\sin \al \sin \theta/2 \end{array}
\right)
e^{i 3\theta/2}
\left( \begin{array}{c} 1 \\ i \\ 0 \end{array} \right).
\eeq
This vortex winds around the singular line part one and a half times
and around the spin part half a time.

One should note that from the above spectrum a
new meaning for the topological charge $\nu$ may be interpreted: as
the winding number of the singular line part of the vortex.   

Another, final, observation that we would like to make is that upon
addition of spatial non-rotationally symmetric terms to the Lagrangian
the only embedded 
vortices that remain solutions to the theory are those which
contain {\em no angular dependence of those spatially associated
components of the order parameter} ({\em i.e.} non are generated by
any part of $SO(3)_L$). With hindsight, this may be expected to be the
case, but it is pleasing to see it coming through in the mathematics.
This leads one to wonder (or conjecture, perhaps) if a similar
phenomena happens in other cases where the spatial rotation group acts
non-trivially upon the order parameter.

\bigskip
\bigskip


{\noindent{\Large{\bf Conclusions}}}

\nopagebreak

\bigskip

\nopagebreak

We conclude by summarising our main results:
\begin{enumerate}
\item In section~(2) we summarised the formalism of the companion paper
  `Embedded Vortices' \cite{me2}. 
\item In section~(3) we rederived the embedded defect spectrum of the
  Weinberg-Salam model. Our results are in agreement with other
  methods. 
\item In section~(4) we derived the embedded defect spectrum of the
  model $SU(3) \rightarrow SU(2)$, finding: embedded monopoles, gauge
  invariant unstable vortices and a family of unstable vortices. 
\item In section~(5) we illustrated `combination vortices' by the model
  $U(1) \x U(1) \rightarrow 1$. This illustrates how such objects may
  only be solutions in certain limits of the coupling constants, and
  the form of their spectrum when such solutions have been found.
\item In section~(6), we examined the embedded defect spectrum
  for three realistic GUT models, namely: Georgi-Glashow $SU(5)$;
  Flipped-$SU(5)$; and Pati-Salam $SU^4(4)$. 
\item Finally, in section~(7), we illustrated how our formalism may also
  be used in some condensed matter contexts --- using the specific
  example of vortices in $^3$He-A. This also illustrated
  combination vortices and some of their stability properties. 
\end{enumerate}

\bigskip
\bigskip


{\noindent{\Large{\bf Acknowledgements.}}}

\nopagebreak

\bigskip

\nopagebreak

This work is supported partly by PPARC partly by the European
Commission under the Human Capital and Mobility programme. One of us
(NFL) acknowledges  
EPSRC for a research studentship. We wish to thank T. Kibble, N. Turok,
T. Vachaspati and G. Volovik for interesting discussions related to
this work.



\end{document}